\documentclass{article}
\usepackage{arxiv}
\usepackage[utf8]{inputenc} % allow utf-8 input
\usepackage[T1]{fontenc}    % use 8-bit T1 fonts
\usepackage{hyperref}       % hyperlinks
\usepackage{url}            % simple URL typesetting
\usepackage{booktabs}       % professional-quality tables
\usepackage{amsfonts}       % blackboard math symbols
\usepackage{nicefrac}       % compact symbols for 1/2, etc.
\usepackage{microtype}      % microtypography
\usepackage{lipsum}		% Can be removed after putting your text content
\usepackage{graphicx}
\usepackage{natbib}
\usepackage{doi}

\title{Quadrupolar interaction induced frequency shift of $^{131}Xe$ nuclear spins on the surface of silicon}

\date{September 29, 2021}	% Here you can change the date presented in the paper title
\author{ \href{https://orcid.org/0000-0002-3169-9577}{\includegraphics[scale=0.06]{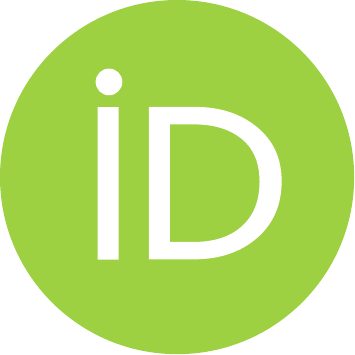}\hspace{1mm}Yao Chen,Mingzhi Yu, Yintao Ma, Libo Zhao, Yanbin Wang, Ju Guo, Qijing Lin and Zhuangde Jiang}\\
	1. School of Mechanical Engineering, Xi'an Jiaotong University, Xi'an 710049, China\\
	2. State Key Laboratory for Manufacturing Systems Engineering,\\ International Joint Laboratory for Micro/Nano Manufacturing and Measurement Technologies,\\ Overseas Expertise Introduction Center for Micro/Nano Manufacturing-\\
	and Nano Measurement Technologies Discipline Innovation,\\ Xi'an Jiaotong University, Xi'an 710049, China\\
    3.Xi'an Jiaotong University Suzhou Institute, Suzhou 215123, China.\\
	\texttt{yaochen@xjtu.edu.cn} \\
		\texttt{libozhao@xjtu.edu.cn} 
}

\renewcommand{\shorttitle}{\textit{arXiv} Template}

\hypersetup{
pdftitle={A template for the arxiv style},
pdfsubject={q-bio.NC, q-bio.QM},
pdfauthor={Yao Chen,Mingzhi Yu, Yintao Ma, Libo Zhao and Zhuangde Jiang},
pdfkeywords={Nuclear Magnetic Resonance Gyroscope, Atomic Co-magnetometer, Atomic Spin Gyroscope,Quadupolar relaxation, Quadrupolar frequency shift},
}

\begin{document}
\maketitle

\begin{abstract}
	The combination of micro-machined technology with the Atomic Spin Gyroscope(ASG) devices could fabricated Chip Scale Atomic Spin Gyroscope(CASG). The core of the gyroscope is a micro-machined vapor cell which contains alkali metal and isotope enriched noble gases such as $^{129}Xe$ and $^{131}Xe$. The quadrupolar frequency shift of $^{131}Xe$ is key parameters which could affect the drift of the ASG and is related to the material of the cell in which they are contained. In micro machined technology, the typical utilized material is silicon. In this article, we studied the electric quadrupolar frequency shift of $^{131}Xe$ atoms with the silicon wall of the micro-machined vapor cell. A cylinder micro-machined vapor cell is utilized in the experiment and a large part of the inner cell surface is composed of silicon material. We studied the temperature dependence of the $^{129}Xe$ spin relaxation and $^{131}Xe$ frequency shifts to evaluate the interaction of the nuclear spin with container wall and the alkali metal atoms. The results show that the average twisted angle of the $^{131}Xe$ nuclear spins as they collide with the silicon wall is measured to be $29 \times 10^{-6} rad$. The desorption energy for the $^{131}Xe$ nuclear spin to escape from the silicon surface is $E_{si}=0.009eV$. This study could help to improve the bias stability of the CASG which is a key parameter for the gyroscope as well as may developes a method to study the surface property of various material.
\end{abstract}

% keywords can be removed

\section{Introduction}
Hyper polarization of isotope enriched nuclear spins\cite{walker1997} could find wide range application, including atomic spin gyroscope\cite{kornack2005,yaochen2016,larsen2012},neutron spin filters\cite{xintong2021},magnetic resonance imaging of the lungs for COVID-19 study\cite{zhouxin2021}, testing physcis beyond the standard model\cite{jiwei2018,lee2018improved,walker2013}, etc. Isotope enriched noble gas nucleus such as $^{131}Xe$,$^{21}Ne$ and $^{83}Kr$ whose nuclear spin are larger than 3/2 own nuclear quadrupole moments. Thus, the study of the nuclear quadrupolar interaction between the nuclear spin and the surrounding enviroment toke attention in several area. For example, in a nuclear magnetic resonance gyroscope or an atomic co-magnetometer\cite{sorensen2020synchronous,xinyexu2021}, the nuclear quadrupolar frequency shift and relaxation of $^{131}Xe$ through colliding with the container wall could affect the bias instability and detection sensitivity of the gyroscope respectively.

The combination of atomic devices with chip-scale fabrication technology could greatly reduce the size and cost of the atomic sensors\cite{kitching2018}. For atomic sensors based on an alkali vapor cell such as atomic clock\cite{knappe2005chip}, atomic magnetometer\cite{knappe2019magnetometer} and atomic spin gyroscope\cite{limesteamless2019}, the key is an alkali vapor cell in which several kinds of gases are filled. Especially in a spin exchange relaxation free(SERF) gyroscope\cite{li2016rotation,duan2018rotation} and nuclear magnetic resonance gyroscope\cite{sorensen2020synchronous}, typically $^{131}Xe$ and $^{21}Ne$ are filled. Traditionally, the glass material is utilized for alkali vapor cell fabrication, while in a micro-machined alkali vapor cell, the typically utilized material is silicon and glass. The glass-silicon anode bonding technology is usually utilized in the cell fabrication process\cite{yaochen2021,runqi2018}. The main body of the cell is composed of a silicon block and two covered glass ends. Ion beam deep etching is usually utilized to drill a hole for the container space of the vapor cell.

Due to the quadrupolar interaction between the $^{131}Xe$ nuclear spins and the vapor cell surface wall, the perturbation from the electric field gradient of the wall surface will cause both the frequency shift and relaxation of the nuclear spins. There are several studies about the nuclear spin-glass colliding\cite{butscher1994,butscher1996nuclear} while there is little study about the nuclear spin-silicon colliding. The frequency shift and relaxation of the nuclear spins relate to the temperature of the vapor cell since the desorption time of the nuclear spins on the surface would affect the nuclear quadrupolar interaction strength\cite{butscher1994}. The desorption is thermally activated and there is a parameter named activation energy $E_A$ which is 0.12eV for the nuclear-glass surface interaction\cite{butscher1994}. Moreover, the shape of the vapor cell is also related to the quadrupolar interaction. We can control the temperature, material and shapes of the vapor cell to change the quadrupolar interaction strength.  For example, in order to study the crossover between the NMR and the nuclear quadrupole resonance interaction regimes, alkali vapor cells with a different kind of material are utilized to make the nuclear quadrupolar frequency shifts more clearly\cite{eadonley2009}. A rectangular alkali vapor cell is also utilized to change the quadrupolar interaction strength\cite{shengdong2020}.

The nuclear quadrupole resonance(NQR) spectroscopy could also be utilized to identify chemicals and it is sometimes called the fingerprint of the chemicals. Thus, the NQR spectroscopy could be utilized to detect explosives\cite{cooper2016}. In this paper, we mainly focus on the NQR spectroscopy study in ASG application. Especially in micro-machined ASGs, the alkali vapor cell' material is composed of both glass and silicon. The quadrupolar interaction between the nuclear spins and the silicon surface is studied. We measured the temperature dependence of the $^{131}Xe$ frequency shift. The NMRG bias instability is close related to the NQR spectroscopy. Thus, this study could help to give solutions to improve the bias instability of the chip scale ASG. 

\section{Theory}
\label{sec:theory}
The quadrupolar interaction happens as the nuclear spin collides with the container wall. The atoms will be absorbed on the surface for a while. This process will cause both the relaxation and frequency shift of the $^{131}Xe$ nuclear spin. The nuclear spin of $^{131}Xe$ atoms is 3/2 and there is a nuclear quadrupole moment in the nucleus. The energy level of the nuclear spin should shift if the nuclear quadrupole moment feels Electric Field Gradient(EFG) as well as the relaxation of the nuclear spin could occur if EFG fluctuation exists. 

The vapor cell utilized in our paper is made of both silicon and glass. As shown in Fig.\ref{fig:cellconfig}, The main body of the vapor cell is made of silicon and the geometry is cylindrical. The inner diameter is 3mm and the length is 2mm in our experiment. The two ends of the vapor cell is covered by glass and it is connected to the silicon through anode bonding. The details about how to fabricate the vapor cell could be found in our paper\cite{yaochen2021}. As the polarized nuclear spin of $^{131}Xe$ atoms colliding with the cell walls, they will be adsorbed by the cell wall for a short period as well as diffuse from site to site during the adsorbed period\cite{butscher1994}. The mean adsorption time $\tau_s$ is defined to be the average time that the atoms are adsorbed on the surface of the cell wall. The adsorption time is related to the temperature of the cell and the time should be decreases if the temperature of cell rise. There is a parameter named $E_A$ which is defined to be the activation energy of desorption to characterize $\tau_s$. The relation is $1/\tau_s \propto exp(-E_A/k_BT)$\cite{butscher1994}. Since both of the frequency shift and relaxation due to quadrupole interaction with cell wall are related to the desorption energy, we can do a measurement of the dependence of frequency shift and relaxation with the cell temperature.

According to this reference\cite{wu1988coherent}, there is a good experiment evidence that the magnitudes of the fluctuating field gradients at the wall are quite large compared to the mean value of the field gradients, and then we assume that:
\begin{equation}
	\left< \left( \frac{\partial^2 V_w}{\partial x_i \partial x_j} \right) ^2 \right> \gg \left< \frac{\partial^2 V_w}{\partial x_i \partial x_j} \right> ^2
\end{equation}
where $V_w$ is the electric field potential at the wall surface, $x_i$ and $x_j$ are unit vectors in the $i$ and $j$ directions. $i$ and $j$ could be $x$,$y$ or $z$ directions. When the atoms are adsorpted on the glass surface wall of the cell, it is plausible to make an assumption that the fluctuations are nearly isotropic, i.e, we assume that the microscopic structure of the wall is sufficiently rough that any tensor components of the electric field gradient have approximately the same mean-squared amplitude as any other. We believe that the silicon surface is also rough and the EFG and its fluctuation are isotropic too. While every material has a characterize deactivation energy. It is reported that in a vapor cell with RbH surface coating, $^{131}Xe$ atoms would stay for a shorter time than that of the glass surface. Thus, the relaxation time of $^{131}Xe$ atoms could be longer\cite{wu1990experimental}. In our experiment, we divided the vapor cell into two parts. One is the glass part and the other one is the silicon part. Both of the two parts can cause the nuclear quadrupole splitting of the energy level and relaxation. According to the reference\cite{wu1988coherent}, the NQR shifts for the $|-3/2><-1/2|$ and $|3/2><1/2|$ coherence as:
\begin{equation}
\label{equatiomega}
	\Delta \Omega=\Delta \Omega_{g}+\Delta \Omega_{si}=\pm \frac{v S}{2V}\frac{1}{2I-1}\left[\int_{S_g}\frac{dS_g}{S}\left< \theta_g \right>\left(\frac{3}{2}cos^2\psi-\frac{1}{2}\right) +\int_{S_{si}}\frac{dS_{si}}{S}\left< \theta_{si} \right>\left(\frac{3}{2}cos^2\psi-\frac{1}{2}\right)\right]
\end{equation}
where $\Delta \Omega_{g}$ and $\Delta \Omega_{si}$ are the frequency shifts from collision with glass and silicon surfaces respectively. $v$ is the velocity of the atoms, $S$ is the inner surface area of the vapor cell and $V$ is the inner volume of the vapor cell. $I$ is the nuclear spin of the atoms. $\left< \theta_g \right>$ and $\left< \theta_{si} \right>$ are the average angles as the nuclear spin colliding with the glass surface and silicon surface respectively. $\psi$ is the angle between the holding magnetic field and the normal direction to the inner surface of the vapor cell. For example, if the holding magnetic field is directed to the $z$ direction and there is a small area $dS_{si}$ on the silicon surface, the angle between the holding magnetic field and the normal direction which is perpendicular to the small area is 90 degree. 

\section{Experimental Setup}
\label{sec:experiment}
\begin{figure}
	\centering
\includegraphics[width=10cm,height=9cm]{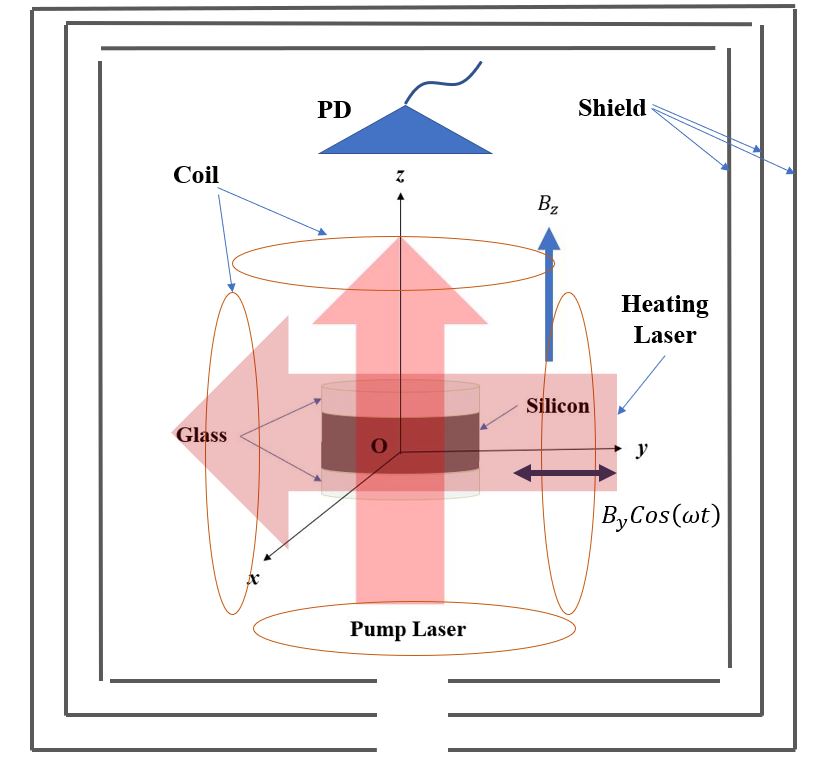}
	\caption{The configuration of the vapor cell, pump laser and the holding magnetic field.}
	\label{fig:cellconfig}
\end{figure}

The configuration of the experimental setup is shown in Fig.\ref{fig:cellconfig}. The micro-machined alkali vapor cell is at the center of the experiment and the geometry is cylinder. The inner diameter is 3mm and the height is 2mm. A small amount of cesium metal, 5 Torr natural abundance Xe gas and 650 Torr Nitrogen gas are filled in the vapor cell. The natural abundance Xe gas contains 26.4\% $^{129}Xe$ and 21.2\% $^{131}Xe$. The pump laser is circular polarized and tuned to the Cs D1 line absorption center. The beam is expanded to cover the cell as large as possible. A second 3W 1550nm heating laser is utilized to heat the vapor cell to the desired temperature. After passing through the vapor cell, a PD(photo diode) is used to accept the transmitted pump laser light. 3 sets of magnetic field coil are utilized for the holding magnetic field in the $z$ direction, the modulation magnetic field $B_y Cos(\omega t)$ in the $y$ direction and the compensation magnetic field in all the three directions. Several layers of magnetic field shields together with a MnZn ferrite shield are utilized for shielding the vapor cell from the earth 's magnetic field. The Xe nuclear spins are hyper-polarized through spin exchange optical pumping with Cs atomic spins. The hyper-polarized Xe nuclear spin will produce magnetic field which could be experienced by the Cs electron spins. The effective magnetic field will be approximately $B^K=8/3\pi k_0 \mu_K [N] P^K$\cite{romalis1998}. In the equation, $B_K$ is the magnetic field produced by the nuclear spin such as $^{129}Xe$ or $^{131}Xe$. $k_0$ is an enhancement factor\cite{walker1989} which could enhance the magnetic field experienced by the Cs electron spins during spin exchange collision. $\mu_K$ is the nuclear magnetic moment for spin species K. $[N]$ is the number density of the nuclear spins and $P^K$ is the polarization of the nuclear spin species K. $^{129}Xe$ nuclear magnetic moment  $\mu(^{129}Xe)$ is -0.78$\mu_N$ in which $\mu_N$ is the nuclear magnetic moment of the neutron and $^{131}Xe$ nuclear magnetic moment $\mu(^{131}Xe)$ is 0.69$\mu_N$. The magnetic field produced by $^{129}Xe$ and $^{131}Xe$ will be in the opposite directions. The Cs atom spins, the pump laser and the modulation magnetic field $B_y Cos(\omega t)$ compose of a single beam absorption magnetometer\cite{shah2007,shah2009} to detect the hyper-polarized Xe nuclear spin magnetic field. The modulation frequency of the magnetic field is 1000Hz which is much larger than the nuclear spin precession frequency.

Since the vapor cell is quite small, it is hard to use free space laser beam for the experiment. A polarization maintaining fiber is utilized for guiding the pump laser to the vapor cell. A multi-mode fiber with 400$\mu m$ core is utilized for the heating laser. The temperature of the vapor cell is measured by the Cs D1 absorption line. We first do a measurement of the line-width of the D1 absorption line. then the Cs atom density could be calculated by the absorption line. Thus we can get the temperature of the vapor cell through the measured number density. We also measured the temperature of the vapor cell through a temperature sensor. The temperature measured by the absorption method is around 5 degree larger than that of the temperature sensor. Here we will use the absorption method, since it directly measures the number density of the Cs atoms.
\begin{figure}
	\centering
\includegraphics[width=16cm,height=6cm]{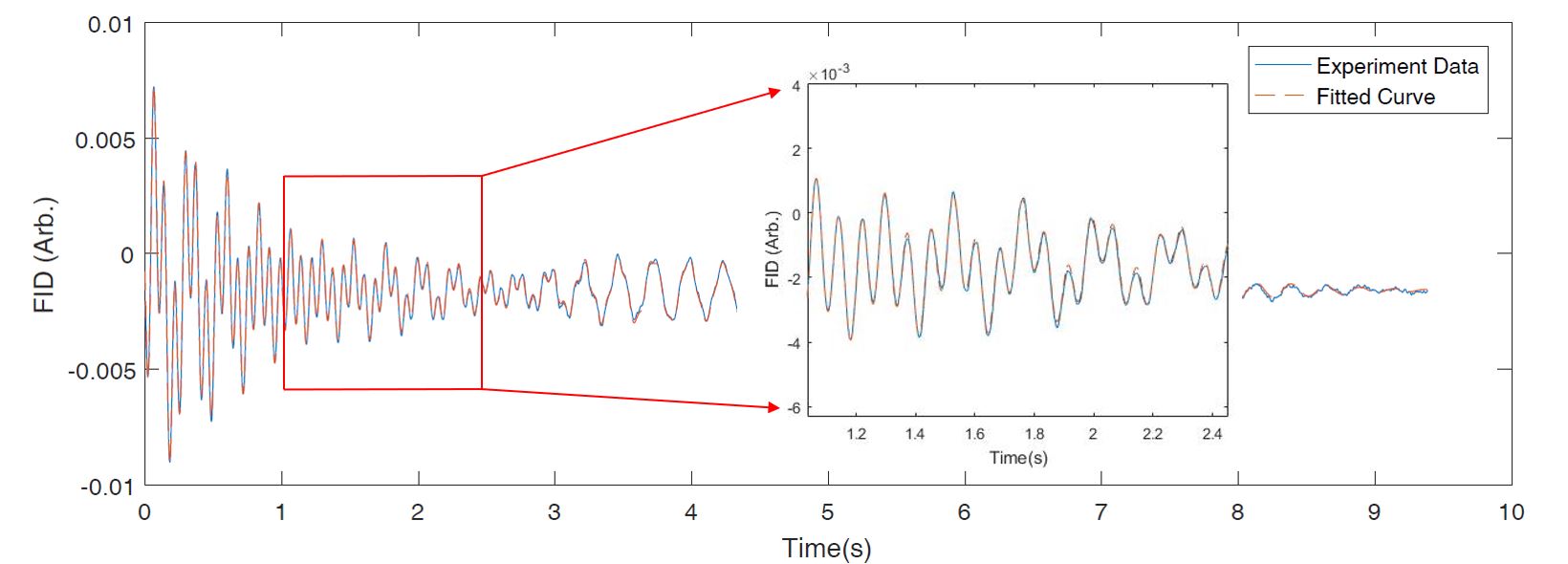}
	\caption{The free induction decay(FID) signal of the $^{129}Xe$ and $^{131}Xe$ nuclear spins. The solid curve is the experiment data and the dashed line is the fitted curve. In order to see the fitting more clearly, we selected a part of the FID signal and amplified it in the embodied figure.}
	\label{fig:129xe131xerelaxation}
\end{figure}

The free induction decay(FID) signal
is a typical way to measure the precession frequency and relaxation of the nuclear spins. Fig.\ref{fig:129xe131xerelaxation} shows one of the FID signal. A holding magnetic field in the $z$ direction $B_z$ is added to the system. After several minutes' spin exchange optical pumping, The polarization tends to be stable. We suddenly add a step magnetic field which is around $1/10$ of the holding magnetic field in the $y$ direction to let the nuclear spin precess around the total magnetic field. After around 1 second, we removed the $y$ step magnetic field and the nuclear spins will precess around the holding magnetic field $B_z$. Note that the residual magnetic field in the shield is below 5nT and we compensated this residual field by the coils with the single beam Cs atomic magnetometer. We mention that the modulation amplitude of the magnetometer only affect the relaxation of Cs atoms. We set the modulation amplitude as small as possible to do the measurement.

\begin{figure}
	\centering
\includegraphics[width=17cm,height=5cm]{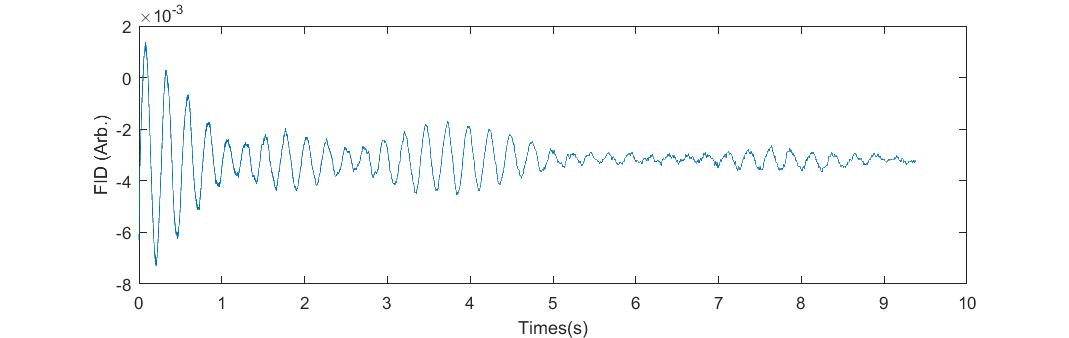}
	\caption{The pure $^{131}Xe$ FID signal which is subtracted from the total signal shown in Figure.\ref{fig:129xe131xerelaxation} by $^{129}Xe$ FID signal.}
	\label{fig:amplitude131}
\end{figure}

The FID signal composes several frequencies. We did a fft analyse of the signal shown in Fig.\ref{fig:129xe131xerelaxation}. We can see that there are 4 frequencies in the signal. The precession frequency of $^{129}Xe$ is single and it is around 3 times of the $^{131}Xe$'s precession frequencies. There are 3 components in the  $^{131}Xe$'s precession. Due to the quadrupolar interaction, the frequency difference of the 3 peaks are the same and equal to $\Delta \Omega$. From Fig.\ref{fig:129xe131xerelaxation} we can see that the high frequency precession is from $^{129}Xe$. There is also the $^{131}Xe$ precession frequency with smaller frequency. The beating signal of $^{131}Xe$ is caused by the nuclear quadrupolar interaction. In order to acquire the frequencies and the relaxation times of the 4 precession components, we do a fitting to the FID experiment data with the following equation:
\begin{tiny}
\begin{equation}
\label{equation3}
	V(t)=AExp(-\Gamma_{129}t)cos(2\pi f_{129}t+\phi_{129})+Exp(-\Gamma_{131}t) \left[ B_1 cos(2\pi f_{131} t+\Delta \Omega t+\phi_{131}^1)+B_2 cos(2\pi f_{131} t+\phi_{131}^2)+B_3 cos(2\pi f_{131} t-\Delta \Omega t+\phi_{131}^3)\right]
\end{equation}
\end{tiny}
where $A$, $B_1$,$B_2$ and $B_3$ are the amplitudes of the 4 precession components. $\Gamma_{129}$ and $\Gamma_{131}$ are the decay rates of the nuclear spin. $f_{129}$ and $f_{131}$ are the precession frequencies of $^{129}Xe$ and $^{131}Xe$ without the nuclear quadrupole shift. $\phi_{129}$, $\phi_{131}^1$, $\phi_{131}^2$ and $\phi_{131}^3$ are the phase of the 4 components respectively. From the fitting result in Fig.\ref{fig:129xe131xerelaxation} we can see that the equation could fit well to the experiment data. In order to see the pure $^{131}Xe$ precession signal, we substract the signal of $^{129}Xe$ from the FID signal. Fig.\ref{fig:amplitude131} shows the pure signal. We can clearly see the beating signal of $^{131}Xe$ precession.
\begin{figure}
	\centering
\includegraphics[width=10cm,height=6cm]{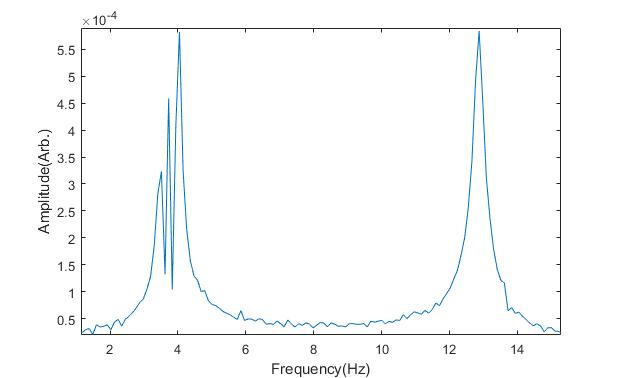}
	\caption{The FFT spectrum of the Xe FID signal.}
	\label{fig:amplitude}
\end{figure}
\section{Results}
\label{sec:results}
\begin{figure}
	\centering
\includegraphics[width=9cm,height=5cm]{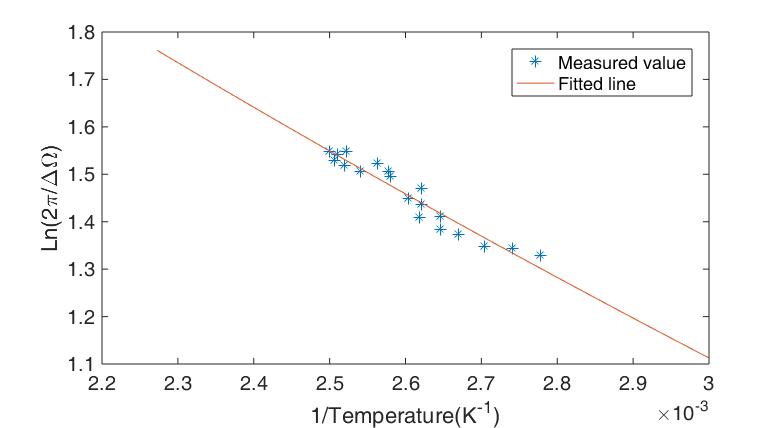}
	\caption{The relationship between the cell temperature and the quadrupole frequency shift of the $^{131}Xe$ energy level.}
	\label{fig:131shift}
\end{figure}

The nuclear quadrupolar shift is related to the temperature of the vapor cell since the temperature could affect $\tau_s$. Finally, the average angle $\left< \theta_g \right>$ could be affected by the temperature. We changed the vapor cell temperature and then measure the $^{131}Xe$ quadrupolar shifts. The relationship between the temperature and the frequency shifts are shown in Fig. \ref{fig:131shift}. According to Equation. \ref{equatiomega}, we can simplify the equation into:
\begin{equation}
\label{equatiomegasim}
	\Delta \Omega=\Delta \Omega_{g}+\Delta \Omega_{si}=\pm \frac{v S}{2V}\frac{1}{2I-1}\frac{d \left< \theta_g \right>-h \left< \theta_{si} \right>}{d+2h}
\end{equation}
where $d$ is the diameter of the vapor cell and $h$ is the height of the vapor cell. Since Ln(1/$\left< \theta \right>) \propto -E_A/(k_B T)$, we set the horizontal axis to be $1/T$ and the vertical axis to be $Ln(2\pi /\Delta \Omega)$. As shown in this reference\cite{butscher1994}, the average angle $\left< \theta_g \right>$ for Pyrex glass is 45$\mu rad$ as the temperature of the vapor cell is 373K. From the fitting result of Fig.\ref{fig:131shift}, as the temperature of the vapor cell is 373K, we can calculate that the average angle for the silicon is 29$\mu rad$ based on equation.\ref{equatiomegasim}. This result is similar to the result in this reference\cite{eadonley2009} in which the angle for silicon is 29$\mu rad$. 

We can also get the desorption energy of $^{131}Xe$ atoms on the surface of silicon $E_{si}$. With the increasing of the temperature, the time that nuclear spins stay on the material will be shorter. Thus, the average angle will be reduced and finally the quadrupolar frequency shift will decrease. In equation.\ref{equatiomegasim}, it is reasonable to suppose that the average angle $\left< \theta \right>$ is propotional to $Exp(E/k_b T)$ in which $E$ is the desorption energy of the material\cite{butscher1994}. Suppose that there are coefficients $k_1$ and $k_2$ which connect the average angle and the desorption energy for the glass and silicon. It is reasonable to set the average thermal velocity of the nuclear spin to be 245$m/s$ which is the velocity under 373K since the thermal velocity is weakly temperature dependent. At 373K, the average angle for the glass is 45$\mu rad$ and the desorption energy is $E_g=0.12eV$\cite{butscher1994}. We can calculate that $k_1$ is equal to $1.1\times10^{-6}$. Together with the experimental condition, we can get:
\begin{equation}
\label{equatiofit}
	Ln\left(\frac{2\pi}{\Delta \Omega}\right)=4.53-\frac{1391.3 x}{c+e x}
\end{equation}
where $c$ is $Ln(606061k_2)$. $x$ is $1/T$ and $e$ is $E_{si}/k_b$. We fit the experimental data shown in Fig.\ref{fig:131shift} with equation\ref{equatiofit}. The fitting results show that $e$ is 109 and $c$ is 0.89. We can get that $E_{si}$ is 0.009 and $k_1$ is equal to $4.0\times10^{-6}$. From the results we see that as the nuclear spin absorbed on the silicon surface, they seems to stay much shorter time than that of the glass. It seems that the EFG is larger on the surface of silicon since the factor $k_2$ is around 3 times larger than that of the glass.   

\begin{figure}
	\centering
\includegraphics[width=8cm,height=5cm]{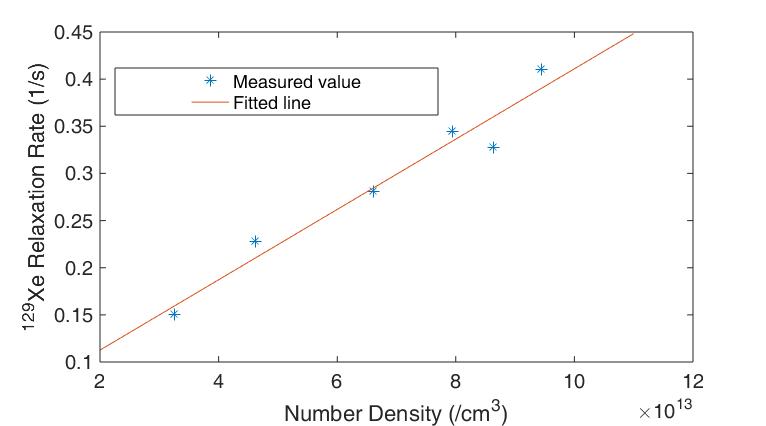}
	\caption{The relationship between the number density of Cs atoms and the relaxation rate of $^{129}Xe$ under low pumping power.}
	\label{fig:129binaryrate}
\end{figure}

Except the frequency shifts, we also studied the relaxation of the $^{129}Xe$ nuclear spins. We changed the number density of the Cs atoms and then measure the relaxation rate of the nuclear spins.   The relaxation of $^{129}Xe$ is also measured through the FID method. The FID signal is fitted to Equation.\ref{equation3} and then the relaxation rate could be measured. The relationship between the Cs number density and $^{129}Xe$ relaxation rate is shown in Fig.\ref{fig:129binaryrate}. The fitting shows that the slope is $3.7 \times 10^{-15} cm^3/s$ and the relaxation rate of $^{129}Xe$ nuclear spin tends to be 0.038 $s^{-1}$ as the Cs number density is 0. Thus the wall relaxation rate of $^{129}Xe$ is 0.038.
During the measurement of the relaxation rate, we find that the relaxation is strongly related to Cs polarization. This is because the polarized Cs atoms could produce magnetic field which could be experienced by the nuclear spins. As the precession frequencies of the nuclear spins and the Cs electron spins close to each other, Cs atom spins would damp the Xe precession\cite{fang2016}. Thus, we lowered the pump laser power as low as possible to do the experiment. We also let the holding magnetic field direct to the nuclear spin magnetic field direction, thus the electron spin of Cs atom could experience both of the nuclear spin magnetic field and the holding magnetic field. This arrangement could set the electron spin precession frequency far away from the nuclear spin precession frequency. During the experiment, we found that the electron spin magnetic field could reach around 200nT at high power. 

The collision between Cs and $^{129}Xe$ could also cause the relaxation. Especially for the heave atoms, binary collision and van der Waals molecule formation could happen at the same time. The interaction time could be long as the molecule forms. Thus nitrogen gas could be filled in the vapor cell as the high density nitrogen molecule could break the Cs-$^{129}Xe$ van der Waals molecules\cite{shao2005measurement}. According to the reference, the binary collision coefficient is $9.4\times 10^{-16}cm^3/s$. For the vapor cell utilized in our experiment, there is 0.85 Amagat nitrogen gas filled. From the measured total relaxation(both of the binary collision and three body collision) result we measured, we can get that the van der Waals molecule formation induced relaxation coefficient is $2.8 \times 10^{-15} cm^3/s$. This result shows that the relaxation induced by the three body collision is about 3 times larger than that of the binary collision relaxation rate.

The relaxation of $^{131}Xe$ is mainly composed of two parts. The first one is the collision relaxation with Cs atoms. This relaxation process should be similar to that of the $^{129}Xe$ atoms. We define this relaxation rate to be $\Gamma ^{131}_{se}$ and it includes the binary spin exchange relaxation and tree body collision relaxation. We define a spin exchange rate coefficient $k_{se}$ and $\Gamma ^{131}_{se}$ is equal to $k_{se} [Cs]$ where $[Cs]$ is the number density of Cs atoms. The other one is the relaxation from the quadrupolar interaction. According to the reference\cite{wu1988coherent}, the quadrupolar relaxation of $^{131}Xe$ for the $|3/2><1/2|$ and $|-3/2><-1/2|$ energy levels are determined to be:
\begin{equation}
\label{equatio131rel}
	\Gamma ^{131}_{Qu}=\frac{2}{5} \frac{v S}{4V} \left(\frac{2h}{d+2h} \left < \theta_{si}^2 \right >+\frac{d}{d+2h} \left < \theta_{g}^2 \right > \right)
\end{equation}
In the equation, $\left < \theta^2 \right >$ is the average squared angle as the nuclear spin collide with the surface wall. We believe that this angle is different for the silicon and glass material. It is also reasonable again to suppose that the EFG fluctuation which could cause the relaxation is isotropic. The coefficients beside the square twisted angle is the inner surface area percentage of the two material. According to the reference\cite{butscher1994}, the average of the squared twisted angle $\left < \theta^2 \right >$ is propotional to $Exp(2E/k_B T)$. Thus, equation.\ref{equatio131rel} could be written:
\begin{equation}
\label{equatio7}
	\Gamma ^{131}_{Qu}=\frac{3}{5} \frac{v S}{4V} \left[\frac{2h}{d+2h} k_1 Exp(\frac{2E_{si}}{k_B T})+\frac{d}{d+2h} k_2 Exp(\frac{2E_{g}}{k_B T}) \right]
\end{equation}
According to the measured results for $E_g=0.12eV$ and $\left < \theta_g^2 \right >=3.4 \times 10^{-6} rad^2$, under the temperature of 373K, we can calculate $k_2$ to be $1.95\times 10^{-9}$ with the relation $\left < \theta_g^2 \right>= k_2 Exp(\frac{2E_g}{k_B T})$. We substitute these results into equation.\ref{equatio7}, the quadrupolar relaxation of $^{131}Xe$ nuclear spin could be:
\begin{equation}
\label{equatio8}
	\Gamma ^{131}_{Qu}=0.125+49100  k_1 Exp(\frac{2E_{si}}{k_B T})
\end{equation}
Now we turn to do a calculation of $\left < \theta_g^2 \right>$. As the cell temperature is 393.5K, the number density of Cs is $6.6\times 10^{13} cm^3$. The total relaxation of $^{131}Xe$ which includes the quadrupolar relaxation with surface and the spin collision with Cs atoms is $0.21 sec^{-1}$. First, we need to substract the relaxation from collision with Cs atoms. There is rare study about the spin exchange collision relaxation between Cs and $^{131}Xe$ atoms. We can only do an empirical calculation of $Cs-^{131}Xe$ spin exchange relaxation from the result of $Cs-^{129}Xe$. Since the angular momentum of $^{131}Xe$ is 3/2 which is 3 times larger than that of the $^{131}Xe$. While the nuclear magnetic moments of the two isotopes are nearly the same. Thus it is reasonable to assume that the spin exchange rate coefficient $k_{se}$ for $Cs-^{129}Xe$ is 3 times that of the $Cs-^{131}Xe$ pair. The measured result for $Cs-^{129}Xe$ pair is $3.7 \times 10^{-15} cm^3/s$ and thus we can calculate the spin exchange rate coefficient $k_{se}^{131}$ to be $1.2 \times 10^{-15} cm^3/s$. Under the temperature of 393.5K, the spin exchange relaxation of $^{131}Xe$ is 0.08$s^{-1}$. We substract the spin exchange relaxation from the total relaxation and then we can get the quadrupolar relaxation of $^{131}Xe$ under 393.5K to be 0.13$s^{-1}$. Since the squared average angle for collision with the glass material is temperature dependent, we can calculate the angle under 393.5K to be $2.3\times 10^{-6} rad^2$. According to equation\ref{equatio131rel}, we can calculate that the glass material induced quadrupolar relaxation to be 0.06$s^{-1}$ under 393.5K. The silicon material induced quadrupolar relaxation is 0.07$s^{-1}$. Finally, we can calculate that $\left < \theta_{si}^2 \right>$ under 393.5K is equal to $2.2\times 10^{-6} rad^2$ which is similar to that of the glass colliding.
\section{Discussion}
\label{sec:discussion}
The glass material utilized in our vapor cell is boro-silicate glass with a thermal expansion coefficient of $3.3\times 10^{-6}/K$. The limitation of this study is that we could not independently measure the quadrupolar interaction between $^{131}Xe$ and the glass surface. We took the parameters from other references for the desorption energy, the mean twisted angle and mean squared twisted angle. Since there is no data about the spin exchange rate between Cs and $^{131}Xe$ nuclear spins, we theoretically estimate this parameter based on the measured rate of $^{129}Xe$. Further studies could be done to directly measure the spin exchange optical pumping rate between Cs and $^{131}Xe$. For comparison, the spin exchange parameters between Cs and $^{129}Xe$ are well known. The other limitation of this study is that we could not measure the relaxation of $^{131}Xe$ in a wide range of temperature. At low temperature and high temperature, the $^{131}Xe$ FID signal will be very weak compared to the $^{129}Xe$ FID signal. Moreover, the FID signal is not so strong since the natural abundance Xe is utilized in our experiment. Other isotopes will contribute to a large part of the Cs spin relaxation and thus the sensitivity of the single beam atomic magnetometer will be low. Further studies could be done to just fill isotope enriched $^{131}Xe$ atoms in the vapor cell. Note that when we fabricating a micro-machined alkali vapor cell, the anode bonding chamber is quite large compared to the glass blown vapor cell fabricating system. It will waste a lot of expensive isotope enriched $^{131}Xe$ for making the micro-machined vapor cell. We are now developing a gas recycling system to improve the situation.
\section{Conclusion}
In conclusion, we have studied the quadrupolar frequency shift and relaxation of $^{131}Xe$ nuclear spins as they collide with the silicon container wall. The silicon material is a kind of widely utilized material in the MEMS technology. This study helps to know more about the surface property of the silicon. CSAG would be developed and this study will finally helps to improve the bias instability of the CSAGs. We divided the alkali vapor cell into the glass parts and the silicon part. Models for the frequency shift and relaxation were developed to measure the quadrupolar interaction between the nuclear spins and the surfaces. The desorption energy of $^{131}Xe$ on the silicon surface is measured to be 0.009eV. At 373K, the average twisted angle $\left < \theta_{g} \right>$ as $^{131}Xe$ collide with the silicon surface wall is 29 $\mu rad$. The relaxation of $^{131}Xe$ was also studied and we acquired the average square twisted angle $\left < \theta_{si}^2 \right>$ which could decide the relaxation of $^{131}Xe$ nuclear spin to be $2.2\times 10^{-6} rad^2$ under 393.5K.
\section{Acknowledgement}
This work is supported by Open Research Projects of Zhejiang Lab under grant number 2019MB0AB02, China Postdoctoral Science Foundation under grant number 2020M683462, National Natural Science Foundation of China under grant number 62103324 and Natural Science Foundation of Jiangsu under grant number BK20200244.
\bibliographystyle{unsrtnat}
\bibliography{references}  %%% Uncomment this line and comment out the ``thebibliography'' section below to use the external .bib file (using bibtex) .

\begin{thebibliography}{32}
\providecommand{\natexlab}[1]{#1}
\providecommand{\url}[1]{\texttt{#1}}
\expandafter\ifx\csname urlstyle\endcsname\relax
  \providecommand{\doi}[1]{doi: #1}\else
  \providecommand{\doi}{doi: \begingroup \urlstyle{rm}\Url}\fi

\bibitem[Walker and Happer(1997)]{walker1997}
Thad~G Walker and William Happer.
\newblock Spin-exchange optical pumping of noble-gas nuclei.
\newblock \emph{Reviews of Modern Physics}, 69\penalty0 (2):\penalty0 629,
  1997.

\bibitem[Kornack et~al.(2005)Kornack, Ghosh, and Romalis]{kornack2005}
TW~Kornack, RK~Ghosh, and Michael~V Romalis.
\newblock Nuclear spin gyroscope based on an atomic comagnetometer.
\newblock \emph{Physical review letters}, 95\penalty0 (23):\penalty0 230801,
  2005.

\bibitem[Chen et~al.(2016)Chen, Quan, Zou, Lu, Duan, Li, Zhang, Ding, and
  Fang]{yaochen2016}
Yao Chen, Wei Quan, Sheng Zou, Yan Lu, Lihong Duan, Yang Li, Hong Zhang, Ming
  Ding, and Jiancheng Fang.
\newblock Spin exchange broadening of magnetic resonance lines in a
  high-sensitivity rotating k-rb-21ne co-magnetometer.
\newblock \emph{Scientific reports}, 6:\penalty0 36547, 2016.
\newblock ISSN 2045-2322.

\bibitem[Larsen and Bulatowicz(2012)]{larsen2012}
Michael Larsen and Michael Bulatowicz.
\newblock Nuclear magnetic resonance gyroscope: For darpa's micro-technology
  for positioning, navigation and timing program.
\newblock In \emph{2012 IEEE International Frequency Control Symposium
  Proceedings}, pages 1--5. IEEE, 2012.
\newblock ISBN 1457718200.

\bibitem[Qin et~al.(2021)Qin, Huang, Buck, Kreuzpaintner, Amir, Salman, Ye,
  Zhang, Jiang, Wang, et~al.]{xintong2021}
Zecong Qin, Chuyi Huang, ZN~Buck, W~Kreuzpaintner, SM~Amir, A~Salman, Fan Ye,
  Junpei Zhang, Chenyang Jiang, Tianhao Wang, et~al.
\newblock Development of a 3he gas filling station at the china spallation
  neutron source.
\newblock \emph{Chinese Physics Letters}, 38\penalty0 (5):\penalty0 052801,
  2021.

\bibitem[Li et~al.(2016{\natexlab{a}})Li, Zhang, Zhao, Sun, Ye, and
  Zhou]{zhouxin2021}
Haidong Li, Zhiying Zhang, Xiuchao Zhao, Xianping Sun, Chaohui Ye, and Xin
  Zhou.
\newblock Quantitative evaluation of radiation-induced lung injury with
  hyperpolarized xenon magnetic resonance.
\newblock \emph{Magnetic resonance in medicine}, 76\penalty0 (2):\penalty0
  408--416, 2016{\natexlab{a}}.

\bibitem[Ji et~al.(2018)Ji, Chen, Fu, Ding, Fang, Xiao, Wei, and
  Yan]{jiwei2018}
Wei Ji, Yao Chen, Changbo Fu, Ming Ding, Jiancheng Fang, Zhigang Xiao, Kai Wei,
  and Haiyang Yan.
\newblock New experimental limits on exotic spin-spin-velocity-dependent
  interactions by using smco5 spin sources.
\newblock \emph{Physical review letters}, 121\penalty0 (26):\penalty0 261803,
  2018.

\bibitem[Lee et~al.(2018)Lee, Almasi, and Romalis]{lee2018improved}
Junyi Lee, Attaallah Almasi, and Michael Romalis.
\newblock Improved limits on spin-mass interactions.
\newblock \emph{Physical review letters}, 120\penalty0 (16):\penalty0 161801,
  2018.

\bibitem[Bulatowicz et~al.(2013)Bulatowicz, Griffith, Larsen, Mirijanian, Fu,
  Smith, Snow, Yan, and Walker]{walker2013}
M~Bulatowicz, R~Griffith, M~Larsen, J~Mirijanian, CB~Fu, E~Smith, WM~Snow,
  H~Yan, and TG~%J Physical review~letters Walker.
\newblock Laboratory search for a long-range t-odd, p-odd interaction from
  axionlike particles using dual-species nuclear magnetic resonance with
  polarized xe 129 and xe 131 gas.
\newblock \emph{Physical review letters}, 111\penalty0 (10):\penalty0 102001,
  2013.

\bibitem[Sorensen et~al.(2020)Sorensen, Thrasher, and
  Walker]{sorensen2020synchronous}
Susan~S Sorensen, Daniel~A Thrasher, and Thad~G Walker.
\newblock A synchronous spin-exchange optically pumped nmr-gyroscope.
\newblock \emph{Applied Sciences}, 10\penalty0 (20):\penalty0 7099, 2020.

\bibitem[Xu et~al.(2021)Xu, Zhou, Peng, Li, Qiu, Zhou, and Xu]{xinyexu2021}
Zhengyi Xu, Yinmin Zhou, Xinxin Peng, Lianhua Li, Xuyang Qiu, Min Zhou, and
  Xinye Xu.
\newblock Measuring the enhancement factor of the hyperpolarized xe in nuclear
  magnetic resonance gyroscopes.
\newblock \emph{Physical Review A}, 103\penalty0 (2):\penalty0 023114, 2021.

\bibitem[Kitching(2018)]{kitching2018}
John Kitching.
\newblock Chip-scale atomic devices.
\newblock \emph{Applied Physics Reviews}, 5\penalty0 (3):\penalty0 031302,
  2018.
\newblock ISSN 1931-9401.

\bibitem[Knappe et~al.(2005)Knappe, Schwindt, Shah, Hollberg, Kitching, Liew,
  and Moreland]{knappe2005chip}
Svenja Knappe, PDD Schwindt, V~Shah, Leo Hollberg, John Kitching, L~Liew, and
  John Moreland.
\newblock A chip-scale atomic clock based on 87 rb with improved frequency
  stability.
\newblock \emph{Optics express}, 13\penalty0 (4):\penalty0 1249--1253, 2005.

\bibitem[Krzyzewski et~al.(2019)Krzyzewski, Perry, Gerginov, and
  Knappe]{knappe2019magnetometer}
SP~Krzyzewski, AR~Perry, V~Gerginov, and S~Knappe.
\newblock Characterization of noise sources in a microfabricated single-beam
  zero-field optically-pumped magnetometer.
\newblock \emph{Biomedical optics express}, 126\penalty0 (4):\penalty0 044504,
  2019.
\newblock ISSN 0021-8979.

\bibitem[Limes et~al.(2019)Limes, Dural, Romalis, Foley, Kornack, Nelson,
  Grisham, and Vaara]{limesteamless2019}
ME~Limes, N~Dural, Michael~V Romalis, EL~Foley, TW~Kornack, A~Nelson,
  LR~Grisham, and J~Vaara.
\newblock Dipolar and scalar he3-xe129 frequency shifts in stemless cells.
\newblock \emph{Physical Review A}, 100\penalty0 (1):\penalty0 010501, 2019.

\bibitem[Li et~al.(2016{\natexlab{b}})Li, Fan, Jiang, Duan, Quan, and
  Fang]{li2016rotation}
Rujie Li, Wenfeng Fan, Liwei Jiang, Lihong Duan, Wei Quan, and Jiancheng Fang.
\newblock Rotation sensing using a k-rb-ne 21 comagnetometer.
\newblock \emph{Physical Review A}, 94\penalty0 (3):\penalty0 032109,
  2016{\natexlab{b}}.

\bibitem[Duan et~al.(2018)Duan, Quan, Chen, Jiang, Fan, Ding, Wang, and
  Fang]{duan2018rotation}
Lihong Duan, Wei Quan, Yao Chen, Liwei Jiang, Wenfeng Fan, Ming Ding, Zhuo
  Wang, and Jiancheng Fang.
\newblock Rotation sensing decoupling of a dual-axis k-rb-21 ne atomic
  comagnetometer.
\newblock \emph{Applied optics}, 57\penalty0 (7):\penalty0 1611--1616, 2018.

\bibitem[Chen et~al.(2021)Chen, Yu, Ma, Luo, Jiang, Bai, and Zhao]{yaochen2021}
Yao Chen, Mingzhi Yu, Yintao Ma, Guoxi Luo, Zhuangde Jiang, Yu~Bai, and Libo
  Zhao.
\newblock Micro-fabricated alkali vapor cells for atomic spin gyroscope study.
\newblock In \emph{2021 IEEE 16th International Conference on Nano/Micro
  Engineered and Molecular Systems (NEMS)}, pages 282--285, 2021.
\newblock \doi{10.1109/NEMS51815.2021.9451404}.

\bibitem[Han et~al.(2018)Han, You, Zhang, Xue, and Ruan]{runqi2018}
Runqi Han, Zheng You, Fan Zhang, Hongbo Xue, and Yong Ruan.
\newblock Microfabricated vapor cells with reflective sidewalls for chip scale
  atomic sensors.
\newblock \emph{Micromachines}, 9\penalty0 (4):\penalty0 175, 2018.

\bibitem[Butscher et~al.(1994)Butscher, Wäckerle, and Mehring]{butscher1994}
R~Butscher, G~Wäckerle, and M~Mehring.
\newblock Nuclear quadrupole interaction of highly polarized gas phase 131xe
  with a glass surface.
\newblock \emph{J The Journal of chemical physics}, 100\penalty0 (9):\penalty0
  6923--6933, 1994.
\newblock ISSN 0021-9606.

\bibitem[Butscher et~al.(1996)Butscher, W{\"a}ckerle, and
  Mehring]{butscher1996nuclear}
R~Butscher, G~W{\"a}ckerle, and M~Mehring.
\newblock Nuclear quadrupole surface interaction of gas phase 83kr: comparison
  with 131xe.
\newblock \emph{Chemical Physics Letters}, 249\penalty0 (5-6):\penalty0
  444--450, 1996.

\bibitem[Donley et~al.(2009)Donley, Long, Liebisch, Hodby, Fisher, and
  Kitching]{eadonley2009}
E.~A. Donley, J.~L. Long, T.~C. Liebisch, E.~R. Hodby, T.~A. Fisher, and
  J.~Kitching.
\newblock Nuclear quadrupole resonances in compact vapor cells: The crossover
  between the nmr and the nuclear quadrupole resonance interaction regimes.
\newblock \emph{Phys. Rev. A}, 79:\penalty0 013420, Jan 2009.
\newblock \doi{10.1103/PhysRevA.79.013420}.
\newblock URL \url{https://link.aps.org/doi/10.1103/PhysRevA.79.013420}.

\bibitem[Feng et~al.(2020)Feng, Zhang, Lu, and Sheng]{shengdong2020}
Y.-K. Feng, S.-B. Zhang, Z.-T. Lu, and D.~Sheng.
\newblock Electric quadrupole shifts of the precession frequencies of
  $^{131}\mathrm{Xe}$ atoms in rectangular cells.
\newblock \emph{Phys. Rev. A}, 102:\penalty0 043109, Oct 2020.
\newblock \doi{10.1103/PhysRevA.102.043109}.
\newblock URL \url{https://link.aps.org/doi/10.1103/PhysRevA.102.043109}.

\bibitem[Cooper et~al.(2016)Cooper, Prescott, Matz, Sauer, Dural, Romalis,
  Foley, Kornack, Monti, and Okamitsu]{cooper2016}
Robert~J. Cooper, David~W. Prescott, Peter Matz, Karen~L. Sauer, Nezih Dural,
  Michael~V. Romalis, Elizabeth~L. Foley, Thomas~W. Kornack, Mark Monti, and
  Jeffrey Okamitsu.
\newblock Atomic magnetometer multisensor array for rf interference mitigation
  and unshielded detection of nuclear quadrupole resonance.
\newblock \emph{Phys. Rev. Applied}, 6:\penalty0 064014, Dec 2016.
\newblock \doi{10.1103/PhysRevApplied.6.064014}.
\newblock URL \url{https://link.aps.org/doi/10.1103/PhysRevApplied.6.064014}.

\bibitem[Wu et~al.(1988)Wu, Schaefer, Cates, and Happer]{wu1988coherent}
Z~Wu, S~Schaefer, GD~Cates, and W~Happer.
\newblock Coherent interactions of the polarized nuclear spins of gaseous atoms
  with the container walls.
\newblock \emph{Physical Review A}, 37\penalty0 (4):\penalty0 1161, 1988.

\bibitem[Wu et~al.(1990)Wu, Happer, Kitano, and Daniels]{wu1990experimental}
Zhen Wu, W~Happer, M~Kitano, and J~Daniels.
\newblock Experimental studies of wall interactions of adsorbed spin-polarized
  xe 131 nuclei.
\newblock \emph{Physical Review A}, 42\penalty0 (5):\penalty0 2774, 1990.

\bibitem[Romalis and Cates(1998)]{romalis1998}
M.~V. Romalis and G.~D. Cates.
\newblock Accurate ${}^{3}\mathrm{He}$ polarimetry using the rb zeeman
  frequency shift due to the $\mathrm{Rb}{\ensuremath{-}}^{3}\mathrm{He}$
  spin-exchange collisions.
\newblock \emph{Phys. Rev. A}, 58:\penalty0 3004--3011, Oct 1998.
\newblock \doi{10.1103/PhysRevA.58.3004}.
\newblock URL \url{https://link.aps.org/doi/10.1103/PhysRevA.58.3004}.

\bibitem[Walker(1989)]{walker1989}
Thad~G. Walker.
\newblock Estimates of spin-exchange parameters for alkali-metal
  \textit{\char21{}} noble-gas pairs.
\newblock \emph{Physical Review A}, 40\penalty0 (9):\penalty0 4959--4964, 1989.
\newblock URL \url{http://link.aps.org/doi/10.1103/PhysRevA.40.4959}.

\bibitem[Shah et~al.(2007)Shah, Knappe, Schwindt, and Kitching]{shah2007}
Vishal Shah, Svenja Knappe, Peter~DD Schwindt, and John Kitching.
\newblock Subpicotesla atomic magnetometry with a microfabricated vapour cell.
\newblock \emph{Nature Photonics}, 1\penalty0 (11):\penalty0 649, 2007.
\newblock ISSN 1749-4893.

\bibitem[Shah and Romalis(2009)]{shah2009}
V~Shah and Michael~V Romalis.
\newblock Spin-exchange relaxation-free magnetometry using elliptically
  polarized light.
\newblock \emph{Physical Review A}, 80\penalty0 (1):\penalty0 013416, 2009.

\bibitem[Fang et~al.(2016)Fang, Chen, Lu, Quan, Zou, and Physics]{fang2016}
Jiancheng Fang, Yao Chen, Yan Lu, Wei Quan, Sheng Zou, and Optical Physics.
\newblock Dynamics of rb and 21ne spin ensembles interacting by spin exchange
  with a high rb magnetic field.
\newblock \emph{Journal of Physics B: Atomic, Molecular}, 49\penalty0
  (13):\penalty0 135002, 2016.
\newblock ISSN 0953-4075.

\bibitem[Shao et~al.(2005)Shao, Wang, and Hughes]{shao2005measurement}
Wenjin Shao, Guodong Wang, and Emlyn~W Hughes.
\newblock Measurement of spin-exchange rate constants between xe 129 and alkali
  metals.
\newblock \emph{Physical Review A}, 72\penalty0 (2):\penalty0 022713, 2005.

\end{thebibliography}

%%% Uncomment this section and comment out the \bibliography{references} line above to use inline references.
% \begin{thebibliography}{1}

% 	\bibitem{kour2014real}
% 	George Kour and Raid Saabne.
% 	\newblock Real-time segmentation of on-line handwritten arabic script.
% 	\newblock In {\em Frontiers in Handwriting Recognition (ICFHR), 2014 14th
% 			International Conference on}, pages 417--422. IEEE, 2014.

% 	\bibitem{kour2014fast}
% 	George Kour and Raid Saabne.
% 	\newblock Fast classification of handwritten on-line arabic characters.
% 	\newblock In {\em Soft Computing and Pattern Recognition (SoCPaR), 2014 6th
% 			International Conference of}, pages 312--318. IEEE, 2014.

% 	\bibitem{hadash2018estimate}
% 	Guy Hadash, Einat Kermany, Boaz Carmeli, Ofer Lavi, George Kour, and Alon
% 	Jacovi.
% 	\newblock Estimate and replace: A novel approach to integrating deep neural
% 	networks with existing applications.
% 	\newblock {\em arXiv preprint arXiv:1804.09028}, 2018.

% \end{thebibliography}

\end{document}